\documentclass[12pt]{iopart}
\usepackage{bm}
\usepackage{graphicx}
\usepackage{hyperref}
\hypersetup{backref=true,pagebackref=true,hyperindex=true,colorlinks=true,breaklinks=true,urlcolor=
red,linkcolor=blue,bookmarks=true,bookmarksopen=true,citecolor=red}

\begin{document}
\title{Spin diffusion and torques in disordered antiferromagnets}
\author{Aurelien Manchon}
\address{Physical Science and Engineering Division (PSE), King Abdullah University of Science and Technology (KAUST), Thuwal 23955-6900,
Kingdom of Saudi Arabia}
\ead{aurelien.manchon@kaust.edu.sa}
\begin{indented}
\item[]June 2016
\end{indented}

\begin{abstract}
We have developed a drift-diffusion equation of spin transport in collinear bipartite metallic antiferromagnets. Starting from a model tight-binding Hamiltonian, we obtain the quantum kinetic equation within Keldysh formalism and expand it to the lowest order in spatial gradient using Wigner expansion method. In the diffusive limit, these equations track the spatio-temporal evolution of the spin accumulations and spin currents on each sublattice of the antiferromagnet. We use these equations to address the nature of spin transfer torque in (i) a spin-valve composed of a ferromagnet and an antiferromagnet, (ii) a metallic bilayer consisting in an antiferromagnet adjacent to a heavy metal possessing spin Hall effect, and in (ii) a single antiferromagnet possessing spin Hall effect. We show that the latter can experience a self-torque thanks to the non-vanishing spin Hall effect in the antiferromagnet.
\end{abstract}
\maketitle

%> start with coupled LLG, show the nature of the torques.
%> start with Hamiltonian, then coupled equations.
%> give the diff equation for an AF
%> give the diff equation for a F for comparison
%> torque in F/AF and AF/AF spin-valves
%> implementing SOC as an oulook
%
%
%Intro: antiferromagnets are coming back; spin Hall and pumping have been observed/predicted;

Antiferromagnets \cite{Neel1932,Neel1970} (AF) have long remained an intriguing and exotic state of matter, whose application has been restricted to enabling interfacial exchange bias \cite{Nogues1999} in metallic and tunneling spin-valves \cite{Dieny1991}. Their role in the expanding field of applied spintronics has been mostly passive and the in-depth investigation of their basic properties considered as fundamental condensed matter physics (see, e.g., \cite{Giamarchi2008,Mourigal2013,Merchant2013}). A conceptual breakthrough was achieved less than ten years ago with the proposal that spin transfer torque could be used to electrically control the direction of the order parameter of antiferromagnetic spin valves, henceforth making these materials potential candidates for low energy spin devices \cite{Nunez2006}. In spite of substantial theoretical efforts and experimental attempts to observe such a torque, the difficulty to independently detect the direction of the antiferromagnetic order parameter has remained a major obstacle. The paradigm has changed radically in the past few years with the discovery of antiferromagnetic anisotropic (tunneling) magnetoresistance \cite{Park2011,Wang2011}, demonstrating that spin-orbit coupled antiferromagnets might emerge as the next frontier in applied spintronics, combining the promises of spin-orbitronics \cite{Manchon2015} and the richness of antiferromagnets \cite{Jungwirth2016,Baltz2016}.

Uncovering the nature of spin torque in antiferromagnetic devices has attracted numerous experimental and theoretical studies. From an experimental standpoint, in spite of an early observation of current-driven change in the exchange bias of a conventional metallic spin-valve \cite{Urazhdin2007,Wei2007}, a major breakthrough has been achieved last year with the observation of current-driven order parameter switching in the non-centrosymmetric antiferromagnet, CuMnAs, mediated by spin-orbit coupling \cite{Wadley2016,Zelezny2014}. Encouraging results have also been observed by Reichlov\'a et al. \cite{Reichlova2015} in Ta/IrMn/CoFeB multilayer stack. From a theoretical perspective, besides a recent derivation of the spin-orbit torque \cite{Zelezny2014}, the physics of spin torque in antiferromagnetic spin devices has been mostly addressed numerically either through tight-binding \cite{Nunez2006,Haney2008,Duine2007,Cheng2012,Prakhya2014,Saidaoui2014,Merodio2015,Saidaoui2016a,Saidaoui2016b} or {\em ab initio} methods \cite{Haney2007,Xu2008}. These works demonstrate without ambiguity that quantum coherence is a crucial ingredient for sizable spin transfer torque in antiferromagnetic spin-valves \cite{Duine2007,Prakhya2014,Saidaoui2014}. The recent realization that the boundary condition at the interface between a normal metal and an antiferromagnet can be modeled through spin mixing conductance \cite{Brataas2000,Brataas2006} enables the phenomenological treatment of interfacial phenomena such as spin torque and spin pumping \cite{Cheng2014,Takei2014}.

Nevertheless, the spin transport in antiferromagnetic devices is still lacking a semiclassical theory that encompasses the important physical parameters governing spin drift and diffusion inside such devices. A drift-diffusion theory would prove very useful to model the spin transport in metallic systems involving antiferromagnets, and thereby complete the drift-diffusion theory already available for ferromagnets \cite{Valet1993,Zhang2002,Petitjean2012}. In this work, starting from a tight-binding Hamiltonian we derive such drift-diffusion equations for a G-type (checkerboard) antiferromagnet and apply the obtained equations to compute the spin transfer torque in three selected cases of interest for experiments:  (i) a spin-valve composed of a ferromagnet and an antiferromagnet, (ii) a metallic bilayer consisting in an antiferromagnet adjacent to a heavy metal possessing spin Hall effect, and in (ii) a single antiferromagnet possessing spin Hall effect.

\section{Antiferromagnetic Dynamics and Torque Definition\label{intro}}
\begin{figure}[h!]
\begin{center}
\includegraphics[width=8cm]{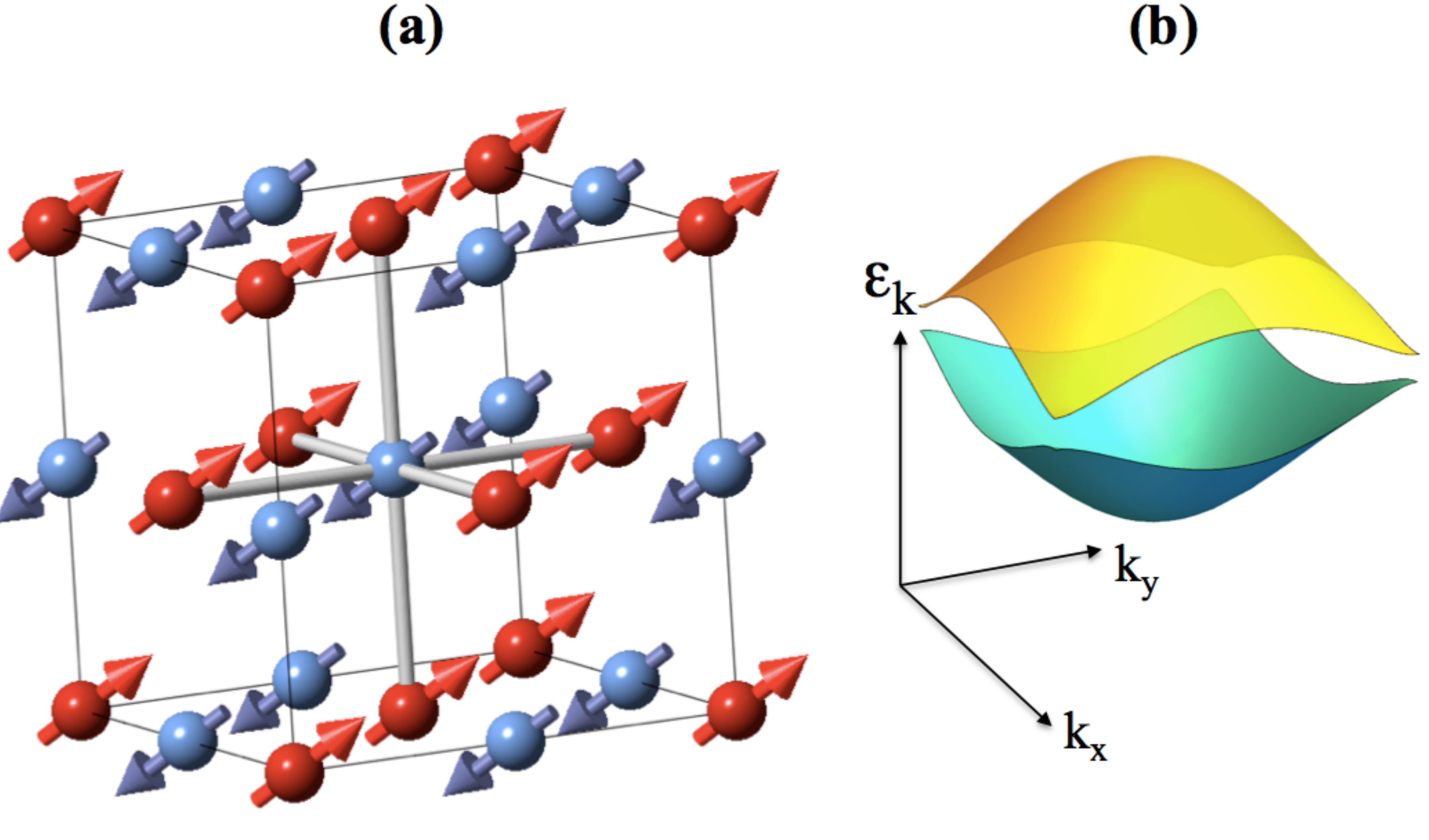}
  \caption{\small{(Color online) (a) Crystal structure of a prototypical G-type antiferromagnet. The red and blue sites and arrows denote the A and B sites and their magnetic moments. The light grey sticks represent the nearest neighbor hopping. (b) Band structure of the G-type antiferromagnet for $k_z=0$.}\label{fig1}}
  \end{center}
\end{figure}

Before deriving the drift-diffusion equation in an antiferromagnet, let us first comment on the nature of the spin transfer torque in these materials. For simplicity, we look at a bipartite antiferromagnet composed of two sublattices A and B, as illustrated in Fig. \ref{fig1}(a). The classical dynamics of such antiferromagnets under a flowing current has already been investigated by Gomonay and Loktev \cite{Gomonay2010,Gomonay2014}, but it is instructive to remind some of the most important aspects of it. Let us consider that each sublattice $i$ is described by Landau-Lifshitz-Gilbert equation
\begin{eqnarray}\fl
\partial_t {\bf m}_i=-\gamma{\bf m}_i\times{\bf H}-\gamma H_{\rm K}({\bf m}_i\cdot{\bf u}){\bf m}_i\times{\bf u}+\gamma H_E{\bf m}_i\times{\bf m}_{\bar{i}}+\alpha {\bf m}_i\times\partial_t{\bf m}_i+{\bm \tau}_i.
\end{eqnarray}
Here ${\bf m}_i$ is the unity vector of the magnetization direction of sublattice $i$, $\gamma$ is the absolute value of the gyromagnetic ratio, $H_{E}$ is the antiferromagnetic exchange field (about 100 T in conventional antiferromagnets), $H_{\rm K}$ is the magnetic anisotropy field along direction ${\bf u}$, $\bar{i}$ is the complementary sublattice and ${\bm \tau}_i$ is the torque exerted on the lattice $i$ by a flowing charge current. In general, it is always possible to express ${\bm \tau}_i$ as
\begin{equation}
{\bm \tau}_i=\tau_\|^i {\bf m}_i\times({\bf p}\times{\bf m}_i)+\tau_\bot^i{\bf p}\times{\bf m}_i,
\end{equation}
where $\tau_{\|,\bot}^i$ is the in-plane (out-of-plane) torque on sublattice $i$ and ${\bf p}$ is a unit vector defined by the spin-polarization mechanism (external polarizer, spin Hall effect, Rashba effect etc.). The antiferromagnet can be characterized by a N\'eel order parameter, ${\bf n}=({\bf m}_A-{\bf m}_B)/2$, and the magnetization ${\bf m}=({\bf m}_A+{\bf m}_B)/2$. Combining the LLG equations for A and B sublattices yields a coupled equation for both ${\bf n}$ and ${\bf m}$. Since the procedure is quite standard and cumbersome, we only provide the final result for the dynamics of ${\bf n}$. This is done by assuming that ${\bf m}$ is a slave variable, an assumption valid for large exchange fields. More precisely, one can show that
\begin{eqnarray}\label{eq:eqm0}\fl
2\gamma H_E{\bf m}&=&\partial_t{\bf n}\times{\bf n}+\gamma{\bf n}\times({\bf H}\times{\bf n})-\frac{1}{2}(\tau_\bot^A+\tau_\bot^B){\bf n}\times({\bf p}\times{\bf n})+\frac{1}{2}(\tau_\|^A-\tau_\|^B){\bf n}\times{\bf p}.
\end{eqnarray}
This equation states that an external magnetic field can induce a magnetization ${\bf m}$ (second term). However, this magnetization remains very small as long as the applied field ${\bf H}$ is smaller than the antiferromagnetic exchange $H_{\rm E}$. Both in-plane and out-of-plane spin transfer torque components can also induce a magnetization as long as their magnitude is {\em staggered} on the sublattices, i.e. if $\tau_\|^A\neq\tau_\|^B$ or $\tau_\bot^A\neq-\tau_\bot^B$. To the first order in magnetization ${\bf m}$, the dynamics equation of the N\'eel order parameter reads
\begin{eqnarray}\label{eq:eqm1}\fl
\partial_t^2{\bf n}\times{\bf n}=-2\gamma^2H_{\rm E}H_{\rm K}({\bf n}\cdot{\bf u}){\bf n}\times{\bf u}+\gamma^2({\bf n}\cdot{\bf H}){\bf n}\times{\bf H}+2\gamma H_{\rm E}\alpha{\bf n}\times\partial_t{\bf n}+2\gamma({\bf n}\cdot{\bf H})\partial_t{\bf n}\nonumber\\
+\gamma H_{\rm E}(\tau_\|^A+\tau_\|^B){\bf n}\times({\bf p}\times{\bf n})+\gamma H_{\rm E}(\tau_\bot^A-\tau_\bot^B){\bf n}\times{\bf p}\nonumber\\
-\frac{1}{2}\partial_t(\tau_\|^A-\tau_\|^B){\bf n}\times{\bf p}+\frac{1}{2}\partial_t(\tau_\bot^A+\tau_\bot^B){\bf n}\times({\bf p}\times{\bf n})-\gamma{\bf n}\times(\partial_t{\bf H}\times{\bf n})\nonumber\\
-\frac{1}{2}(\tau_\|^A-\tau_\|^B)\partial_t{\bf n}\times{\bf p}+\frac{1}{2}(\tau_\bot^A+\tau_\bot^B)\partial_t[({\bf n}\cdot{\bf p}){\bf n}].
\end{eqnarray}
The first line describes the usual dynamics of antiferromagnets \cite{Gomonay2010,Gomonay2014}. The magnetic field only acts on the order parameter at the second order (second term) and through an additional friction term, $\sim\partial_t{\bf n}$ (fourth term). Most importantly, the second line shows that the in-plane and out-of-plane components of the torque induce N\'eel order dynamics  only as long as their magnitude is {\em uniform} on the sublattices. This means that the ability of the spin transfer torque to excite the N\'eel order parameter does not depend on whether the torque is in or out of the $({\bf n},{\bf p})$ plane, but rather on whether the torque magnitude itself is staggered or uniform on the sublattices, an essential feature already noticed in Refs. \cite{Jungwirth2016,Zelezny2014,Gomonay2014}. The third line states although a uniform magnetic field (or staggered spin torque) does not affect the N\'eel order dynamics, its {\em time derivative} can. Such dynamics has been illustrated by the recent demonstration of N\'eel order manipulation using ultrashort laser pulses \cite{Kimel2004,Wienholdt}. Finally, the fourth line shows that staggered torques enhance the energy dissipation of the N\'eel order parameter dynamics, $\sim\partial_t{\bf n}$.

To summarize this section, one should always keep in mind that in antiferromagnets the torques that are efficient in manipulating the N\'eel order parameter must be {\em uniform} on the two sublattices, and therefore arise from {\em staggered} magnetic fields. In contrast, torques that are staggered on the sublattices and arise from uniform magnetic fields can only manipulate the N\'eel order through their time derivative, i.e. using electrical pulses for instance.
\section{Drift-Diffusion Equations in Antiferromagnets}
%%%%%%%%%%%%%%%%%%%%%%%%%%%%%%%%%%%%%%%%%%%%%%%%%%%%%%%%%%%
\subsection{Four band model}
Let us now turn our attention towards spin transport in disordered antiferromagnets. We consider a NaCl crystal lattice composed of two interpenetrating fcc sublattices, A and B, whose magnetic moments are antiferromagnetically coupled and aligned along the direction of the order parameter $\bf n$, as illustrated in Fig. \ref{fig1}(a). In this work, we consider only nearest neighbor hopping and neglect next-nearest neighbor hopping. The extension of the present method to layered antiferromagnets is also possible be remains out of the scope of the present study. The nearest-neighbor tight-binding Hamiltonian in the ($|A\rangle,|B\rangle$)$\otimes$($|\uparrow\rangle,|\downarrow\rangle$) basis reads
\begin{eqnarray}\label{eq:huawei}
\tilde{H}_0&=&\gamma_k\hat\tau_x\otimes\hat{1}+\Delta\hat{\bm\sigma}\cdot{\bf n}\otimes\hat\tau_z,\\
\gamma_k&=&-2t_{\rm N}\sum_{i\neq j}\cos\frac{k_ia}{{2}}\cos\frac{k_ja}{{2}}
\end{eqnarray}
where $\hat{\bm\tau}$ and $\hat{\bm\sigma}$ are the vectors of  2$\times$2 spin Pauli matrices referring to the sublattice and spin angular momentum spaces, respectively, $t_{\rm N}$ is the nearest neighbor hopping integral and $\Delta$ is the exchange energy between the itinerant electrons and the localized electrons. Symbols $\hat{}$ and $\tilde{}$ denote 2$\times$2 and 4$\times$4 operators, respectively. Close to $\Gamma$-point, ${\bf k}\rightarrow 0$ and $\gamma_k\rightarrow ta^2k^2-6t_{\rm N}$. In the presence of spin-independent impurities, we obtain the following Hamiltonian
\begin{eqnarray}\label{eq:h2}
\tilde{H}=&\gamma_k\hat\tau_x\otimes\hat{1}+\Delta\hat{\bm\sigma}\cdot{\bf n}\otimes\hat\tau_z+\tilde{V}({\bf r}).
\end{eqnarray}
The impurity potential consists in short range (delta-like) potentials randomly distributed over the volume of the crystal,
\begin{eqnarray}\label{eq:self}
\tilde{V}({\bf r})=\frac{1}{2}(1+\hat\tau_z)\sum_{i\in \Omega_A} V_0\delta({\bf r}-{\bf r}_i)+\frac{1}{2}(1-\hat\tau_z)\sum_{i\in \Omega_B} V_0\delta({\bf r}-{\bf r}_i),
\end{eqnarray}
where the operators $\frac{1}{2}(1\pm\hat\tau_z)$ refer to the A and B sublattice, respectively, and the summation runs over the volume occupied by each sublattice, $\Omega_{A,B}$. This form ensures that the disorder on sublattice A is not correlated with disorder on sublattice B.\par

The unperturbed retarded (advanced) Green's function associated with $\tilde{H}_0$ is defined as $\tilde{G}^{R,A}_0=(E-\tilde{H}_0\pm i\eta)^{-1}$, and reads explicitly
\begin{eqnarray}\label{eq:go}
\tilde{G}^{R,A}_{k,0}=\frac{1}{2}\sum_{\nu=\pm1}\frac{\tilde{1}+\nu\cos\chi_k\hat\tau_x\otimes\hat{1}+\nu\sin\chi_k\hat{\bm\sigma}\cdot{\bf n}\otimes\hat\tau_z}{\epsilon-\epsilon_{k,\nu}\pm i0^+}
\end{eqnarray}
where 
\begin{eqnarray}
\epsilon_{k,\nu}&=&\nu\sqrt{\gamma_k^2+\Delta^2},\;\cos\chi_k=\frac{\gamma_k}{\sqrt{\gamma_k^2+\Delta^2}},\;\sin\chi_k=\frac{\Delta}{\sqrt{\gamma_k^2+\Delta^2}}.
\end{eqnarray}
With the notation given above, the gap is centered at $\epsilon=0$. From now on, we consider the bottom band ($\nu=-1$) and the electron energy will be taken negative, $-\sqrt{(6t_{\rm N})^2+\Delta^2}<\epsilon<0$. As a result, in the reciprocal space and within first Born approximation, the lesser, retarded and advanced self-energies read
\begin{eqnarray}\label{eq:self0}
\tilde\Sigma^{<,R,A}&=&\int\frac{d^3{\bf k}}{(2\pi)^3}\tilde{V}_k\tilde{G}_k^{<,R,A}\tilde{V}_k,
\end{eqnarray}
$\tilde{V}_k$ being the Fourier transform of $\tilde{V}({\bf r})$. More explicitly
\begin{eqnarray}\label{eq:selfretarded}
\tilde\Sigma^{R,A}&=&\mp\frac{i}{4\tau}(\tilde{1}-\beta\hat{\bm\sigma}\cdot{\bf n}\otimes\hat\tau_z),\\
\tilde\Sigma^{<}&=&\frac{1}{4\pi{\cal N}\tau}(\langle\tilde G_k^<\rangle+\hat\tau_z\langle\tilde G_k^<\rangle\hat\tau_z).\label{eq:selflesser}
\end{eqnarray}
Here, we defined the density of states ${\cal N}=\int d^3{\bf k}/(2\pi)^3\delta(\epsilon-\epsilon_k)$, and the momentum relaxation time $1/\tau=2\pi n_i{\cal N}|V_0^2|$, where $n_i$ is the impurity concentration, and $\beta=\cos\chi_{k_{\rm F}}$ is the onsite polarization at Fermi energy. Finally, $\langle\tilde G_k^<\rangle=\int d^3{\bf k}/(2\pi)^3\tilde G_k^<$.

%%%%%%%%%%%%%%%%%%%%%%%%%%%%%%%%%%%%%%%%%
\subsection{Quantum Kinetics}
Let us now derive the equation of motion of electrons flowing in the antiferromagnet. To do so, we start from Dyson equation,
\begin{eqnarray}
[\hat{\cal G}^R]^{-1}\star\hat{\cal G}^<-\hat{\cal G}^<\star[\hat{\cal G}^A]^{-1}=\hat{\cal S}^<\star\hat{\cal G}^A-\hat{\cal G}^R\star\hat{\cal S}^<,
\end{eqnarray}
where the Green's functions $\tilde{\cal G}=\tilde{\cal G}({\bf r},{\bf r}';t,t')$ and self-energies $\tilde{\cal S}=\tilde{\cal S}({\bf r},{\bf r}';t,t')$ are defined in real space and time, and $\star$ is the convolution product in both space and time. We then perform a Wigner expansion in the center of mass coordinates $(({\bf r}+{\bf r}')/2,(t+t')/2)$, while the short range variations $({\bf r}-{\bf r},t-t')$ are Fourier transformed. This method allows for rewriting Dyson equation in k-space to the lowest order in spacial and temporal gradients, thereby separating the (real space-time) semiclassical dynamics from the (reciprocal space-time) quantum effects. Since this method been described in detailed in several publications \cite{Kovalev2008,Mischenko2004,Wang2012,Shen2014,Ndiaye2015,Ortiz2016}, we directly provide the kinetic equation
%\begin{equation}\fl
%i\hbar\partial_t\tilde G_k^<+[\tilde{G}^<_k,\tilde{H}_0]+i\{\tilde{G}^<_k,\tilde\Sigma\}+\frac{i}{2}\{\hat{v}_i,\partial_i\tilde G^<_k\}+\frac{i}{2}\{eE_i\hat{v}_i,\partial_{\omega}\tilde G^<_{\rm eq}\}+ieE_i\partial_{k_i}\tilde G^<_{\rm eq}=\tilde{\Sigma}^<\tilde{ G}_k^A-\tilde{ G}_k^R\tilde{ \Sigma}^<,\label{qke}
%\end{equation}
\begin{equation}
i\hbar\partial_t\tilde G_k^<+[\tilde{G}^<_k,\tilde{H}_0]+i\{\tilde{G}^<_k,\tilde\Sigma\}+\frac{i}{2}\{\hat{v}_i,\partial_i\tilde G^<_k\}=\tilde{\Sigma}^<\tilde{ G}_k^A-\tilde{ G}_k^R\tilde{ \Sigma}^<,\label{qke}
\end{equation}
where we defined $\hat\Sigma^{R,A}=\mp i\hat\Sigma$, and the velocity operator $\hat{v}_i=\partial_{\hbar k_i}\tilde{H}_0$. In order to obtain sublattice-resolved coupled equations, we now expand the lesser Green's function $\tilde G_k^<$ in the sublattice basis. We write 
\begin{eqnarray}
\tilde G_k^<=\frac{1}{2}(1+\hat\tau_z)\otimes\hat g_{k}^A+\frac{1}{2}(1-\hat\tau_z)\otimes\hat g_{k}^B+\hat\tau_x\otimes\hat g_{k}^x+\hat\tau_y\otimes\hat g_{k}^y.
\end{eqnarray}
Here, $\hat g_{k}^\alpha$ is a 2$\times$2 spinor in the spin basis, $\hat g_{k}^{A,B}$ are the components of the lesser Green's function on sublattice A and B, while $\hat g_{k}^{x,y}$ contain the off-diagonal elements that connect the two sublattices. Hence, by taking the trace of Eq. (\ref{qke}) over the components of $\hat{\bm\tau}$, one obtains four coupled equations
\begin{equation}\fl
\partial_t\hat g_{k}^A-\frac{2\gamma_k}{\hbar}\hat g_k^y-i\frac{\Delta}{\hbar}[\hat g_{k}^A,\hat{\bm\sigma}\cdot{\bf n}]+v_i\partial_i\hat g_k^x=\frac{1}{2\tau}\left(\langle{\hat g}_k^A\rangle\delta_k-{\hat g}_k^A\right)-\frac{\beta}{4\tau}\{\langle{\hat g}_k^A\rangle\delta_k-{\hat g}_k^A,\hat{\bm\sigma}\cdot{\bf n}\},\label{eq:1}
\end{equation}
%\begin{equation}\fl
%\left(\partial_t+\frac{1}{2\tau}\right)\hat g_{k}^B+2\gamma_k\hat g_k^y-i\Delta[\hat g_{k}^B,\hat{\bm\sigma}\cdot{\bf n}]+\frac{\beta}{4\tau}\{\hat g_k^B,\hat{\bm\sigma}\cdot{\bf n}\}+v_i\partial_i\hat g_k^x=\frac{\delta_k}{2{\cal N}\tau}\left(\langle{\hat g}_k^B\rangle+\frac{\beta}{2}\{\langle{\hat g}_k^B\rangle,\hat{\bm\sigma}\cdot{\bf n}\}\right),
%\end{equation}
\begin{equation}\fl
\partial_t\hat g_{k}^B+\frac{2\gamma_k}{\hbar}\hat g_k^y+i\frac{\Delta}{\hbar}[\hat g_{k}^B,\hat{\bm\sigma}\cdot{\bf n}]+v_i\partial_i\hat g_k^x=\frac{1}{2\tau}\left(\langle{\hat g}_k^B\rangle\delta_k-{\hat g}_k^B\right)+\frac{\beta}{4\tau}\{\langle{\hat g}_k^B\rangle\delta_k-{\hat g}_k^B,\hat{\bm\sigma}\cdot{\bf n}\},
\end{equation}
\begin{equation}\fl
\partial_t\hat g_{k}^x+\frac{1}{2\tau}\hat g_{k}^x+\frac{\Delta}{\hbar}\{\hat g_{k}^y,\hat{\bm\sigma}\cdot{\bf n}\}-\frac{i\beta}{4\tau}[\hat g_k^y,\hat{\bm\sigma}\cdot{\bf n}]+\frac{v_i}{2}\partial_i(\hat g_k^A+\hat g_k^B)=0,
\end{equation}
\begin{equation}\fl
\partial_t\hat g_{k}^y+\frac{1}{2\tau}\hat g_{k}^y+\frac{\gamma_k}{\hbar}(\hat g_k^A-\hat g_k^B)-\frac{\Delta}{\hbar}\{\hat g_{k}^x,\hat{\bm\sigma}\cdot{\bf n}\}+\frac{i\beta}{4\tau}[\hat g_k^x,\hat{\bm\sigma}\cdot{\bf n}]=0.\label{eq:2}
\end{equation}
We define $\delta_k=\delta(\epsilon-\epsilon_k)/{\cal N}$. The electric field can be installed by performing the substitution $v_i\partial_i\rightarrow v_i\partial_i-eE_iv_i\partial_\epsilon$. Equations (\ref{eq:1})- (\ref{eq:2}) constitute the semiclassical basis on which the drift-diffusion theory is built.

%%%%%%%%%%%%%%%%%%%%%%%%%%%%%%%%%%%%%%%%%
\subsection{Drift-Diffusion Equations}
The physics emerging out of Eqs. (\ref{eq:1})- (\ref{eq:2}) is difficult to analyze at this stage. We now proceed with momentum averaging in order to obtain the coupled diffusion equations. We define the density matrices on sublattices A and B as $\hat\rho_{A,B}=\int d^3{\bf k}/(2\pi)^3\hat g_k^{A,B}/2\pi{\cal N}$ and the current density spinor as  $\hat{\cal J}_i=\int d^3{\bf k}/(2\pi)^3 v_i\hat g_k^x/\pi{\cal N}$. We also define the auxiliary quantities $\hat\rho_y=\int d^3{\bf k}/(2\pi)^3\hat g_k^y/2\pi{\cal N}$ and $\hat{\cal J}_i^y=\int d^3{\bf k}/(2\pi)^3 v_i\hat g_k^y/2\pi{\cal N}$. These last two quantities do not have a straightforward physical meaning and can be eliminated to derive diffusion equations involving only $\hat \rho_{A,B}$ and $\hat{\cal J}_i$.\par

 After some algebra, one obtains
\begin{equation}\fl
\left(1+\frac{\xi^2}{2}-\frac{\beta^2}{2}\right)\hat{\cal J}_i+\left(\frac{\xi^2}{2}+\frac{\beta^2}{2}\right)\hat{\bm\sigma}\cdot{\bf n}\hat{\cal J}_i\hat{\bm\sigma}\cdot{\bf n}=-{\cal D}\partial_i(\hat\rho_A+\hat\rho_B),
\end{equation}
\begin{equation}\fl
\partial_t\hat\rho_A+\frac{\Gamma}{\hbar}(\hat\rho_A-\hat\rho_B)-i\frac{\Delta}{\hbar}[\hat\rho_A,\hat{\bm\sigma}\cdot{\bf n}]=-\frac{1}{2}\partial_i\hat{\cal J}_i,
\end{equation}
\begin{equation}\fl
\partial_t\hat\rho_B-\frac{\Gamma}{\hbar}(\hat\rho_A-\hat\rho_B)+i\frac{\Delta}{\hbar}[\hat\rho_B,\hat{\bm\sigma}\cdot{\bf n}]=-\frac{1}{2}\partial_i\hat{\cal J}_i.
\end{equation}
Here ${\cal D}=2\tau v_{\rm F}^2/3$ is the diffusion coefficient, $\xi=4\tau\Delta/\hbar$, and $\Gamma=4\tau\epsilon^2(1-\beta^2)$. Finally, by recognizing that $\hat{\cal J}_i=j_{c,i}+\hat{\bm\sigma}\cdot{\bf J}^s_i$, where $j_{c,i}$ is the charge current density and ${\bf J}^s_i$ the spin current density flowing along direction $i$, and $\hat\rho_\alpha=n_\alpha+\hat{\bm\sigma}\cdot{\bf S}_\alpha$, where $n_\alpha$ and ${\bf S}_\alpha$ are the charge and spin densities on sublattice $\alpha$, we obtain
\begin{equation}\label{eq:na}
\partial_tn_A+\frac{\Gamma}{\hbar}(n_A-n_B)=-\frac{1}{2}\partial_ij_{c,i},
\end{equation}
\begin{equation}\label{eq:nb}
\partial_tn_B-\frac{\Gamma}{\hbar}(n_A-n_B)=-\frac{1}{2}\partial_ij_{c,i},
\end{equation}
\begin{equation}\label{eq:Sa}
\partial_t{\bf S}_A+\frac{\Gamma}{\hbar}({\bf S}_A-{\bf S}_B)+\frac{2\Delta}{\hbar}{\bf S}_A\times{\bf n}+\frac{1}{\tau_{\rm sf}}{\bf S}_A=-\frac{1}{2}\partial_i{\bf J}_i^s,
\end{equation}
\begin{equation}\label{eq:Sb}
\partial_t{\bf S}_B-\frac{\Gamma}{\hbar}({\bf S}_A-{\bf S}_B)-\frac{2\Delta}{\hbar}{\bf S}_B\times{\bf n}+\frac{1}{\tau_{\rm sf}}{\bf S}_B=-\frac{1}{2}\partial_i{\bf J}_i^s.
\end{equation}
Equations (\ref{eq:na})-(\ref{eq:Sb}) describes the diffusion of the charge and spin densities on sublattices A and B, in which spin relaxation $\sim 1/\tau_{\rm sf}$ has been introduced by hand. The source of the charge/spin dynamics is given by the gradient of the charge and spin current, i.e. $\partial_ij_{c,i}$ and $\partial_i{\bf J}_i^s$. The charge and spin currents are defined
\begin{equation}
j_{c,i}=-{\cal D}^\|\partial_i (n_A+n_B),
\end{equation}
\begin{equation}\label{eq:Jsaf}
{\bf J}_i^s=-{\cal D}^\|\partial_i [({\bf S}_A+{\bf S}_B)\cdot{\bf n}]{\bf n}-{\cal D}^\bot{\bf n}\times[\partial_i({\bf S}_A+{\bf S}_B)\times{\bf n}],
\end{equation}
In other words, the charge (spin) current $j_{c,i}$ (${\bf J}_i^s$) is the total charge (spin) current flowing through the diatomic unit cell. Here, ${\cal D}^\|={\cal D}/(1+\xi^2)$, ${\cal D}^\bot={\cal D}/(1-\beta^2)$ (Notice that by definition, $|\epsilon|>\Delta$). Most importantly, the charge/spin dynamics on the two sublattices are coupled through a term $\sim \Gamma$. This term is proportional to the ratio between the energy broadening due to disorder ($\sim \hbar/\tau$) and the kinetic energy of the carriers ($\langle \gamma_k\rangle$) and as such, the coupling term $\Gamma$ accounts for the time a spin carrier spends on one sublattice. One can then foresee that there is an interesting interplay between the lifetime of the carrier on a certain sublattice, $\hbar/\Gamma$, and the spin precession time, $\tau_\Delta=\hbar/2\Delta$.\par

This interplay arises explicitly by manipulating Eqs. (\ref{eq:Sa}) and (\ref{eq:Sb}) in a more convenient form. In the limit of slow spin dynamics $(\hbar/\Gamma)\partial_t\ll 1$, we obtain
\begin{equation}\label{eq:naf}
\partial_t(n_A+n_B)=-\partial_i j_{c,i},
\end{equation}
\begin{equation}\label{eq:saf}
\partial_t({\bf S}_A+{\bf S}_B)+\frac{1}{\tau_\varphi}{\bf n}\times[({\bf S}_A+{\bf S}_B)\times{\bf n}]+\frac{1}{\tau_{\rm sf}}({\bf S}_A+{\bf S}_B)=-\partial_i{\bf J}_i^s,
\end{equation}
\begin{equation}\label{eq:stgspin}
{\bf S}_A-{\bf S}_B=\frac{\tau^*}{\tau_\Delta}{\bf n}\times({\bf S}_A+{\bf S}_B),
\end{equation}
where $1/\tau^*=\Gamma/\hbar+1/\tau_{\rm sf}$ and $\tau_\varphi=\tau^*/\tau_\Delta^2$. Equation (\ref{eq:saf}) describes the diffusion of the {\em uniform} spin density ${\bf S}_A+{\bf S}_B$, while Eq. (\ref{eq:stgspin}) defines the {\em staggered} spin density ${\bf S}_A-{\bf S}_B$. In Eq. (\ref{eq:saf}), the third term on the left-hand side, $\sim 1/\tau_{\rm sf}$, relaxes the (uniform) spin density isotropically, and the second term, $\sim 1/\tau_{\varphi}$, relaxes only the component that is transverse to the magnetic order parameter ${\bf n}$. As such, this equation suggests that an antiferromagnet behaves like a normal metal with an {\em anisotropic spin relaxation}. \par

From the spin transfer torque standpoint, the {\em staggered} spin density ${\bf S}_A-{\bf S}_B$ is the most important quantity, as discussed in Section \ref{intro}. The staggered spin density arises from the precession of the uniform spin density about the local order parameter ${\bf n}$. The magnitude of the staggered spin density is governed by the ratio between the carrier lifetime on one sublattice, $\sim\tau^*$, and the spin precession time, $\sim\tau_\Delta$. As a matter of fact, when the mobility is weak and carrier lifetime large compared to the spin precession time ($\tau^*\gg\tau_\Delta$), the incoming spin has the time to precess about the local magnetic moment, thereby generating a staggered spin density. On the contrary, when the carrier lifetime is shorter than the spin precession time ($\tau^*\ll\tau_\Delta$), the itinerant spin jumps quickly from one sublattice to the other without having the time to precess significantly. Therefore, the incoming spin density is weakly affected by the precession and the staggered spin density vanishes.

\section{Spin Torque in Devices Based on Antiferromagnets}

\begin{figure}[h!]
\begin{center}
\includegraphics[width=10cm]{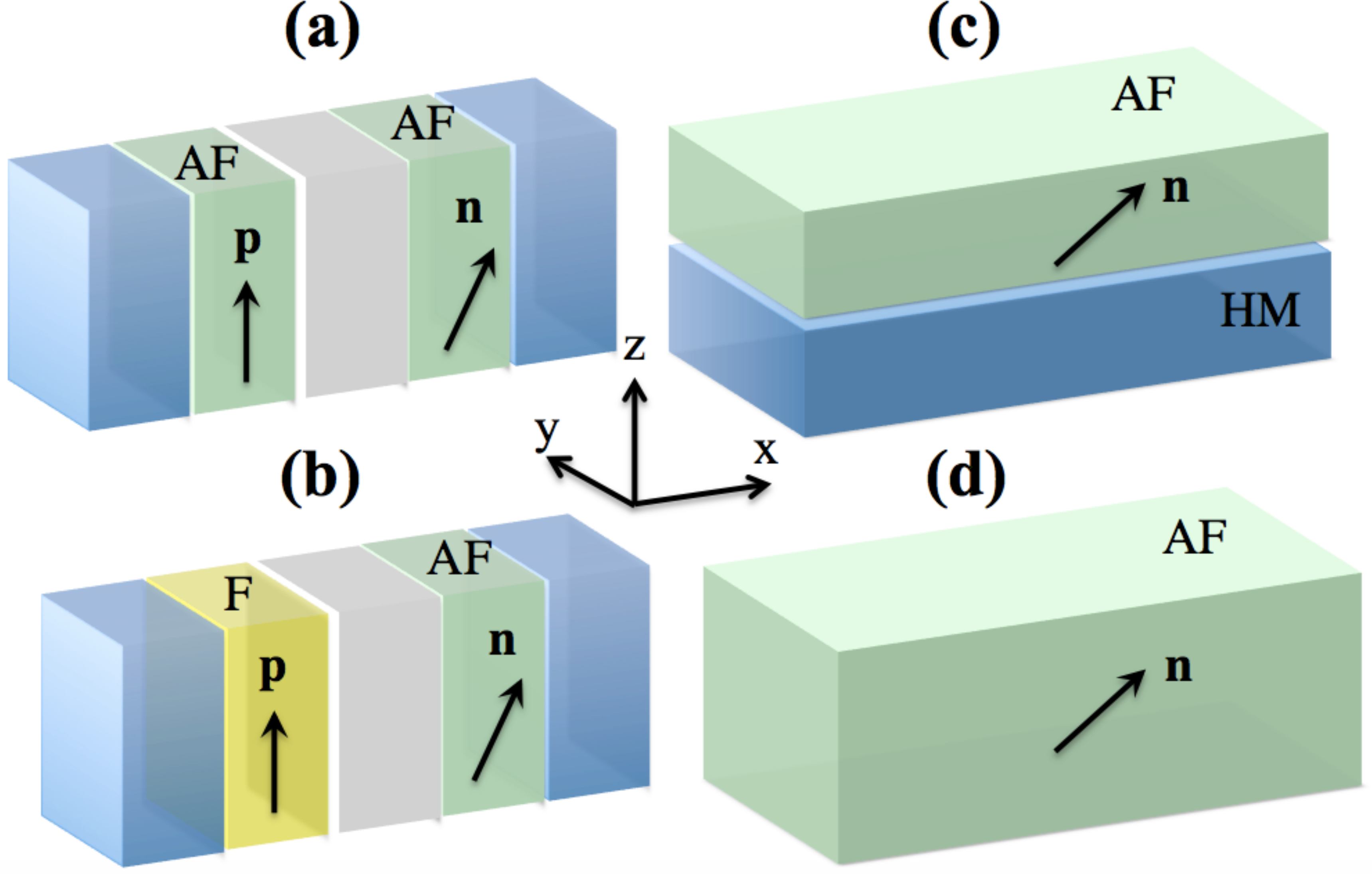}
  \caption{\small{(Color online) Schematics of four spin devices based on antiferromagnets: (a) a spin-valve composed of two antiferromagnets, (b) a spin-valve composed of a ferromagnetic polarizer and an antiferromagnet, (c) a bilayer composed of an antiferromagnet deposited on a heavy metal, and (d) a single antiferromagnet with spin-orbit coupling.}\label{fig2}}
  \end{center}
\end{figure}

Let us now apply the drift-diffusion theory developed in the previous section to four spin devices involving antiferromagnets and illustrated in Fig. \ref{fig2}: (a) a spin-valve composed of two antiferromagnets, (b) a spin-valve composed of a ferromagnetic polarizer and an antiferromagnetic free layer, (c) a bilayer composed of an antiferromagnet deposited on top of a heavy metal and (d) an antiferromagnet with spin-orbit coupling. 

\subsection{Antiferromagnetic/Antiferromagnetic Spin-Valve}

Such a spin-valve has been investigated numerically in several works \cite{Nunez2006,Duine2007,Xu2008,Saidaoui2014} and it is found that the current-driven spin density is dominated by a uniform out-of-plane contribution, with a smaller staggered in-plane component. In other words, using our notation, ${\bf S}_A+{\bf S}_B\sim {\bf n}\times{\bf p}$, and ${\bf S}_A-{\bf S}_B\sim {\bf n}\times({\bf p}\times{\bf n})$. However, these works point out that the non-equilibrium spin density only survive when quantum coherence is preserved  \cite{Duine2007,Saidaoui2014}. In these calculations, the non-equilibrium spin density arises from quantum-coherent reflections at the interfaces between the antiferromagnets. In the diffusive limit, Eq. (\ref{eq:stgspin}), the staggered spin density stems as a {\em correction} of the uniform spin density ${\bf S}_A+{\bf S}_B$. In Eq. (\ref{eq:saf}) the source of uniform spin density is given by the spin current gradient $-\partial_i{\bf J}_i^s\sim \partial_i^2({\bf S}_A+{\bf S}_B)$. In other words, the uniform spin density can only be non-zero if there is a source of spin polarization of some sort (external spin polarizer, spin Hall effect effect etc.). As a result, it is simply not possible to achieve spin transfer torque in a spin-valve composed of two antiferromagnets in the semiclassical limit.

\subsection{Ferromagnetic/Antiferromagnetic Spin-Valve}

An efficient method to exert a torque on an antiferromagnet is to consider a spin-valve composed of a ferromagnetic spin-polarizer and a "free" antiferromagnet \cite{Haney2008}, as depicted in Fig. \ref{fig2}(b). In this model, we consider two semi-infinite (ferro or antiferro)magnetic electrodes separated by a metallic spacer. The antiferromagnet has the usual G-type configuration discussed above, for which spin transport is governed by Eqs. (\ref{eq:saf})-(\ref{eq:stgspin}). To connect our drift-diffusion equations to that in ferromagnets \cite{Zhang2002,Petitjean2012}, we consider that the ferromagnet has a diatomic unit cell similar to the G-type antiferromagnet, with A and B sublattices. In contrast with the antiferromagnet though, in the ferromagnet the magnetic moments on A and B sublattices are ferromagnetically coupled. Hence, the usual spin diffusion equations apply on the total spin density ${\bf S}_A+{\bf S}_B$ of the ferromagnetic unit cell and read (see for instance Refs. \cite{Zhang2002,Petitjean2012})
\begin{equation}\fl\label{eq:Jsf}
{\bf J}_i^s=-{\cal D} \beta {\cal N} eE_i{\bf m}-{\cal D}\partial_i ({\bf S}_A+{\bf S}_B),
\end{equation}
\begin{equation}\fl\label{eq:sf}
\partial_t({\bf S}_A+{\bf S}_B)+\frac{1}{\tau_\Delta}{\bf m}\times({\bf S}_A+{\bf S}_B)+\frac{1}{\tau_\varphi}{\bf m}\times[({\bf S}_A+{\bf S}_B)\times{\bf m}]+\frac{1}{\tau_{\rm sf}}({\bf S}_A+{\bf S}_B)=-\partial_i{\bf J}_i^s,
\end{equation}
Notice that the main difference between the spin current in a ferromagnet, Eq. (\ref{eq:Jsf}), and the spin current in an antiferromagnet, Eq. (\ref{eq:Jsaf}), is that the spin current in the ferromagnet can be directly generated by an electric field [first term in Eq. (\ref{eq:Jsf})] while is it not possible to generate such a spin current in an antiferromagnet. This obvious difference stems from the fact that the density of states in ferromagnets is spin-polarized, contrary to that in antiferromagnets. We also emphasize that Eq. (\ref{eq:sf}) displays a precession term ($\sim 1/\tau_\Delta$) that is absent in antiferromagnets, Eq. (\ref{eq:saf}).\par

We now simply need to connect the three layers. We make the following simplification: we assume that there is no interfacial spin relaxation between the layers (the spin current is continuous), and the spin relaxation in the spacer is much longer than the width of the spacer. We also consider that there is an interfacial resistance $r_{\rm int}$ between the spacer and the antiferromagnet. We then obtain the uniform spin density in the ferromagnet,
\begin{eqnarray}\fl\label{eq:af}
{\bf S}_A+{\bf S}_B|_{x<x_{L}}&=&
\frac{\beta j_c}{e^3{\cal N}}\frac{e^{(x-x_L)/\lambda_\|^L}}{{\rm Den}}\left(\eta_\|\left[\eta_\bot\sigma_\bot^{L}+\sigma_\bot^{R}+\frac{\eta_\bot^2\sigma_\Delta^{ L2}}{\chi_\bot\sigma_\bot^{ R}}\right]-(\eta_\|\sigma_\bot^R-\eta_\bot\sigma_\|^R)\sin^2\theta\right){\bf p}\\
&&+\frac{\beta j_c}{e^3{\cal N}}\frac{e^{(x-x_L)/\lambda_{\bot}^L}}{{\rm Den}}(\eta_\|\sigma_\bot^R-\eta_\bot\sigma_\|^R)\left[\cos\frac{x-x_L}{\lambda_\Delta^L}+\frac{\sigma_\Delta^L\eta_\bot}{\sigma_\Delta^R\chi_\bot}\sin\frac{x-x_L}{\lambda_\Delta^L}\right]\cos\theta{\bf p}\times({\bf n}\times{\bf p})\nonumber\\
&&+\frac{\beta j_c}{e^3{\cal N}}\frac{e^{(x-x_L)/\lambda_{\bot}^L}}{{\rm Den}}(\eta_\|\sigma_\bot^R-\eta_\bot\sigma_\|^R)\left[\sin\frac{x-x_L}{\lambda_\Delta^L}-\frac{\sigma_\Delta^L\eta_\bot}{\sigma_\Delta^R\chi_\bot}\cos\frac{x-x_L}{\lambda_\Delta^L}\right]\cos\theta{\bf p}\times{\bf n},\nonumber
\end{eqnarray}
and in the antiferromagnet,
\begin{eqnarray}\fl\label{eq:f}
{\bf S}_A+{\bf S}_B|_{x>x_{R}}&=&
\frac{\beta j_c}{e^3{\cal N}}\frac{e^{-(x-x_R)/\lambda_\|^R}}{{\rm Den}}\left[\sigma_\bot^{ L}+\sigma_\bot^{ R}+\frac{\eta_\bot^2\sigma_\Delta^{ L2}}{\chi_\bot\sigma_\bot^{ R}}\right]\cos\theta{\bf n}\\
&&+\frac{\beta j_c}{e^3{\cal N}}\frac{e^{-(x-x_R)/\lambda_{\bot}^R}}{{\rm Den}}\left[\eta_\|\sigma_\bot^{ L}+\sigma_\|^{R}+\frac{\eta_\bot\eta_\|\sigma_\Delta^{ L2}}{\chi_\bot\sigma_\bot^{ R}}\right]{\bf n}\times({\bf p}\times{\bf n})\nonumber\\
&&+\frac{\beta j_c}{e^3{\cal N}}\frac{e^{-(x-x_R)/\lambda_{\bot}^R}}{{\rm Den}}\left[\eta_\|\sigma_\bot^{R}-\eta_{\bot}\sigma_\|^{R}\right]\cos\theta{\bf n}\times{\bf p}\nonumber.
\end{eqnarray}
We defined the spin conductivity $\sigma_\nu^\alpha={\cal D}^\alpha/e^2{\cal N}\lambda_\nu^\alpha$ (in units of $\Omega^{-1}\cdot$ m$^{-2}$), where $\nu=\|,\bot,\Delta$, and $\alpha=L,R$ (L stands for the ferromagnet, while R stands for the antiferromagnet). $\lambda_\|^\alpha$ is the relaxation length for spins longitudinal to the magnetic order parameter, while $\lambda_\bot^\alpha$ is the relaxation length for spins transverse to it. $\lambda_\Delta^L$ is the precession length in the ferromagnet. We also note $\eta_\nu=1+r_{\rm int}\sigma_\nu^R$, and $\chi_\bot=1+\eta_\bot\sigma_\bot^L/\sigma_\bot^R$.\par

It is interesting to notice that the uniform spin density injected from the ferromagnet into the antiferromagnet is mostly {\em in-plane} [$\sim {\bf n}\times({\bf p}\times{\bf n})$]. Remarkably, the {\em out-of-plane} component ($\sim {\bf n}\times{\bf p}$) is proportional to $\eta_\|\sigma_\bot^{R}-\eta_{\bot}\sigma_\|^{R}$, in other word, the out-of-plane component of the spin density in the antiferromagnet comes from the anisotropy of the spin relaxation $\lambda_\bot^R\neq\lambda_\|^R$. In turn, this anisotropy induces a spin density in the ferromagnet that is transverse to the magnetization direction [second and third terms in Eq. (\ref{eq:f})]. This means that, in principle, due to the spin relaxation anisotropy in the antiferromagnet, a torque can be exerted from the antiferromagnet on the ferromagnet. Finally, we also note that the components of the spin density that are proportional to $\eta_\|\sigma_\bot^{R}-\eta_{\bot}\sigma_\|^{R}$ are also proportional to $\cos\theta$, i.e. they vanish when the ferromagnetic order parameter is orthogonal to the antiferromagnetic order parameter, a feature already noticed by Haney and MacDonald \cite{Haney2008}.\par

Let us now address the torque exerted by the ferromagnet on the antiferromagnet. By definition, the torque arising from the staggered spin density reads
\begin{eqnarray}
{\bf T}&=&2\Delta\int_{x_R}^{+\infty} dx({\bf S}_A-{\bf S}_B)\times{\bf n}=\frac{\tau^*\hbar}{\tau_\Delta^2}\int_{x_R}^{+\infty} dx{\bf n}\times[({\bf S}_A+{\bf S}_B)\times{\bf n}],
\end{eqnarray}
which yields
\begin{eqnarray}
{\bf T}&=&\lambda_\bot^R\frac{\tau^*\hbar}{\tau_\Delta^2}\frac{\beta j_c}{e^3{\cal N}}\frac{1}{{\rm Den}}\left[\eta_\|\sigma_\bot^{ L}+\sigma_\|^{R}+\frac{\eta_\bot\eta_\|\sigma_\Delta^{ L2}}{\chi_\bot\sigma_\bot^{ R}}\right]{\bf n}\times({\bf p}\times{\bf n})\\
&&+\lambda_\bot^R\frac{\tau^*\hbar}{\tau_\Delta^2}\frac{\beta j_c}{e^3{\cal N}}\frac{1}{{\rm Den}}\left[\eta_\|\sigma_\bot^{R}-\eta_{\bot}\sigma_\|^{R}\right]\cos\theta{\bf n}\times{\bf p}\nonumber.
\end{eqnarray}
Again, the torque is dominated by an in-plane component and possesses a small out-of-plane component, the latter vanishing when the order parameters are orthogonal to each other. In the limit where the spin relaxation is isotropic in the antiferromagnet ($\lambda_\bot^R\approx\lambda_\|^R$), the torque reduces
\begin{eqnarray}\label{eq:torquesv}
{\bf T}&=&\frac{\tau^*\hbar}{\tau_\Delta^2}\frac{1}{e^3{\cal N}}\frac{\lambda_\|^R\beta j_c}{\sigma_\|^R+\sigma_\|^L+r_{\rm int}\sigma_\|^L\sigma_\|^R}{\bf n}\times({\bf p}\times{\bf n}).
\end{eqnarray}
The current-driven dynamics of the in-plane torque has been investigated by Gomonay \cite{Gomonay2010,Gomonay2014} and we refer the reader to these works for further details.

\subsection{Antiferromagnetic bilayer}
In the previous section, we showed that the spin torque arising from a ferromagnetic polarizer in a spin-valve configuration is efficient in manipulating the order parameter of an antiferromagnet. Yet, the fabrication of such a device remains challenging and a much simpler configuration is a magnetic bilayer that consists of an antiferromagnet deposited on top of a heavy metal \cite{Liu2011,Uchida2010}. In the heavy metal, spin-orbit coupling is large enough to enable spin Hall effect. This configuration has been recently investigated experimentally by Reichlov\'a et al. \cite{Reichlova2015}. In the heavy metal, the uniform spin density fulfills the following transport equations \cite{Haney2013,Shchelushkin2005}
\begin{equation}\fl\label{eq:she}
{\bf J}_i^s=-{\cal D}\nabla_i({\bf S}_A+{\bf S}_B)+\alpha_{\rm H}{\cal D}{\bf e}_i\times{\bm\nabla}(n_A+n_B),
\end{equation}
\begin{equation}\fl\label{eq:sf}
\partial_t({\bf S}_A+{\bf S}_B)+\frac{1}{\tau_{\rm sf}}({\bf S}_A+{\bf S}_B)=-\partial_i{\bf J}_i^s,
\end{equation}
where the second term in Eq. (\ref{eq:she}) stands for the spin Hall effect induced by the charge gradient (or equivalently, an electric field). In the configuration depicted on Fig. \ref{fig2}(c), we consider that the electric field is applied along $x$, i.e. ${\bm\nabla}(n_A+n_B)=-{\cal N}eE{\bf x}$, while the normal to the interface is along $z$. The heavy metal has a thickness $d$, while the antiferromagnet is much thicker than its spin relaxation length. We also consider that there is an interfacial resistivity $r_{\rm int}$ between the two layers, at $z=0$. We assume that there is no spin current flowing through the outer boundary of the heavy metal (${\cal J}_z^s|_{z=-d}=0$). Therefore, one simply needs to connect the spin current and densities at the interface between the antiferromagnet and the heavy metal. One obtains
\begin{eqnarray}\fl\label{eq:bilayer}
{\bf S}_A+{\bf S}_B|_{x>0}&=&
\frac{\alpha_{\rm H} j_c}{e^3{\cal N}}\frac{e^{-x/\lambda_\|^R}}{\sigma_\|^R+\eta_\|\sigma_{\rm sf}^L\tanh\frac{d}{\lambda_{\rm sf}^L}}n_y\left(1-\cosh^{-1}\frac{d}{\lambda_{\rm sf}^L}\right){\bf n}\\
&&+\frac{\alpha_{\rm H} j_c}{e^3{\cal N}}\frac{e^{-x/\lambda_\bot^R}}{\sigma_\bot^R+\eta_\bot\sigma_{\rm sf}^L\tanh\frac{d}{\lambda_{\rm sf}^L}}\left(1-\cosh^{-1}\frac{d}{\lambda_{\rm sf}^L}\right){\bf n}\times({\bf p}\times{\bf n})\nonumber.
\end{eqnarray}
One can remark that the uniform spin density in the antiferromagnet does not possess out-of-plane component, in contrast to the spin-valve case studied above. The resulting torque reads
\begin{eqnarray}\fl\label{eq:bilayer}
{\bf T}=\frac{\tau^*\hbar}{\tau_\Delta^2}\frac{1}{e^3{\cal N}}\frac{\lambda_\bot^R\alpha_{\rm H} j_c}{\sigma_\bot^R+\eta_\bot\sigma_{\rm sf}^L\tanh\frac{d}{\lambda_{\rm sf}^L}}\left(1-\cosh^{-1}\frac{d}{\lambda_{\rm sf}^L}\right){\bf n}\times({\bf p}\times{\bf n}).
\end{eqnarray}
The structure of this expression is very similar to Eq. (\ref{eq:torquesv}), which comes as no surprise since the only distinction between the spin-valve and the bilayer structures comes from the nature of the spin polarization (either from a ferromagnetic polarizer or from spin Hall effect).

\subsection{Self-Torque in Single Antiferromagnets}

Let us now turn our attention towards the fourth case represented on Fig. \ref{fig2}, the homogeneous antiferromagnet. We now solve Eq. (\ref{eq:saf}) by assuming that spin Hall effect is present in the antiferromagnet. Spin Hall effect does not modify the spin diffusion equation, but rather the spin current definition. Since an antiferromagnet behaves essentially like a normal metal, it is sufficient to expend the definition of the spin current, Eq. (\ref{eq:Jsaf}), by accounting for spin Hall effect
\begin{equation}\fl
{\bf J}_i^s=-{\cal D}\nabla_i({\bf S}_A+{\bf S}_B)+\alpha_{\rm H}{\cal D}{\bf e}_i\times{\bm\nabla}(n_A+n_B),
\end{equation}
where we neglected the anisotropy of the diffusion coefficients for simplicity. To ensure that an effective torque applies on the system, one has to make the boundary conditions asymmetric otherwise, no effective torque survive after averaging out over the sample volume. For instance, we assume a thickness $d$, such that at $z=0$ the spin density vanishes, while at $z=d$ the spin current vanishes. Therefore, one the uniform spin density reads
\begin{equation}
{\bf S}_A+{\bf S}_B=\frac{1}{e^3{\cal N}}\frac{\alpha_{\rm H}j_c}{\sigma_\bot}\frac{\sinh\frac{z}{\lambda_\bot}}{\cosh\frac{z}{\lambda_\bot}}{\bf n}\times({\bf y}\times{\bf n}),
\end{equation}
and the associated torque
\begin{equation}
{\bf T}=\frac{\tau^*\hbar}{\tau_\Delta^2}\frac{1}{e^3{\cal N}}\frac{\lambda_\bot\alpha_{\rm H}j_c}{\sigma_\bot}\left(1-\cosh^{-1}\frac{d}{\lambda_\bot}\right){\bf n}\times({\bf y}\times{\bf n}).
\end{equation}
Of course the exact expression depends on the exact boundary conditions at $z=0$ and $z=d$. Experimentally, the symmetry can be broken by imposing strong spin-flip at one interface (by dusting the surface with heavy metal impurities for instance) and weak spin-flip at the other interface (using either a light metal like Cu or a specular tunnel barrier like MgO). Finally, noticing that $\lambda_\bot/e^2{\cal N}\sigma_\bot=\tau_\bot$ is the transverse spin relaxation time, we notice that the three torques derived in this work have the same general expression:
\begin{equation}
{\rm Polarization}\cdot\frac{{\rm Life\;time}}{{\rm Precession\;time}}\cdot\frac{{\rm Transverse\;relaxation\;time}}{{\rm Precession\;time}}.
\end{equation}

\section{Conclusion}
We have developed a drift-diffusion theory of spin transport in collinear bipartite metallic antiferromagnets. Using this equation, we derived an expression of the spin transfer torque exerted on the antiferromagnetic free layer in three experimentally-relevant configurations: (i) a spin-valve composed of a ferromagnet and an antiferromagnet, (ii) a metallic bilayer consisting in an antiferromagnet adjacent to a heavy metal possessing spin Hall effect, and in (ii) a single antiferromagnet possessing spin Hall effect. We show that in all three cases, the uniform torque arising from the staggered spin density lies in the plane, $\sim {\bf n}\times({\bf p}\times{\bf n})$.\par
This work shows that spin transfer torque in spin devices involving antiferromagnets survives in the diffusive regime, which is quite promising for potential applications. Several issues remain to be solved though \cite{Jungwirth2016,Baltz2016}. The electrical detection of the N\'eel order dynamics is probably the most significant challenge, although recent breakthroughs exploiting anisotropic magnetoresistance are encouraging \cite{Wadley2016}. A second problem is the quality of the interface between antiferromagnets and normal metals and in particular the nature of the antiferromagnetic spin texture at this interface \cite{Nogues1999}. Finally, the present work is limited to the archetypal collinear G-type antiferromagnet whereas the most common metallic antiferromagnets, such as IrMn, possess a non-collinear antiferromagnetic texture. Extending the present model to such non-collinear antiferromagnets and understanding their current-induced dynamics remain to be explored carefully.

\section{Acknowledgments}
This work was supported by the King Abdullah University of Science and Technology (KAUST) through the Office of
Sponsored Research (OSR) [Grant Number OSR-2015-CRG4-2626]. The author acknowledges inspiring discussions with T. Jungwirth, J. Sinova, J. Zelezny and H. Saidaoui.

\end{document}